\documentclass[manuscript,screen]{acmart}

\AtBeginDocument{%
  \providecommand\BibTeX{{%
    \normalfont B\kern-0.5em{\scshape i\kern-0.25em b}\kern-0.8em\TeX}}}

\setcopyright{acmcopyright}
\copyrightyear{2025}
\acmYear{2025}
\acmDOI{XXXXXXX.XXXXXXX}

\usepackage{multirow}	
\usepackage{tabularx}	
\usepackage{xcolor}	
\usepackage{url}	
\usepackage{xspace}	
\usepackage{caption}	
\usepackage{subcaption}	
\usepackage{float}	
\usepackage{hyperref}

\usepackage{cleveref}	
\usepackage{float}	
\hypersetup{	
    colorlinks=true,	
    linkcolor=blue,	
    filecolor=magenta,      	
    urlcolor=cyan,	
}
\usepackage{hyperxmp}
\usepackage{enumitem}	
\begin{document}

%_____________________ COMMENTS  _____________________ (toggle submittrue/submitfalse)	
\definecolor{Author1}{HTML}{e41a1c}  %% red
\definecolor{Author2}{HTML}{377eb8}  %% blue
\definecolor{Author3}{HTML}{4daf4a}  %% green
\definecolor{Author4}{HTML}{984ea3}  %% Purple
\definecolor{Author5}{HTML}{ff7f00}  %% orange
% For revision	
% \definecolor{Issue1}{HTML}{e41a1c}  %% red
% \definecolor{Issue2}{HTML}{377eb8}  %% blue
% \definecolor{Issue3}{HTML}{4daf4a}  %% green
% \definecolor{Issue4}{HTML}{984ea3}  %% Purple
% \definecolor{Issue5}{HTML}{ff7f00}  %% orange
\setlist{leftmargin=3mm}
\newif\ifsubmit	

\submittrue	
% \submitfalse

\ifsubmit
\newcommand{\zinat}[1]{}	
\newcommand{\zinatIn}[1]{}	
\newcommand{\ray}[1]{}	
\newcommand{\rayIn}[1]{}	
\newcommand{\vivian}[1]{}	
\newcommand{\vivianIn}[1]{}	

%Issue-wise commands	

\else
%Author-wise commands
\newcommand{\zinat}[1]{\marginpar{\colorbox{Author1}{\textcolor{white}{ZA}} \textcolor{Author1}{#1}}}
\newcommand{\zinatIn}[1]{\colorbox{Author1}{\textcolor{white}{ZA}} \textcolor{Author1}{#1}}
\newcommand{\ray}[1]{\marginpar{\colorbox{Author2}{\textcolor{white}{RH}} \textcolor{Author2}{#1}}}
\newcommand{\rayIn}[1]{\colorbox{Author2}{\textcolor{white}{RH}} \textcolor{Author2}{#1}}
\newcommand{\steve}[1]{\marginpar{\colorbox{Author3}{\textcolor{white}{SP}} \textcolor{Author3}{#1}}}
\newcommand{\steveIn}[1]{\colorbox{Author3}{\textcolor{white}{SP}} \textcolor{Author3}{#1}}
\newcommand{\amanda}[1]{\marginpar{\colorbox{Author4}{\textcolor{white}{AH}} \textcolor{Author4}{#1}}}
\newcommand{\amandaIn}[1]{\colorbox{Author4}{\textcolor{white}{AH}} \textcolor{Author4}{#1}}
\newcommand{\hemant}[1]{\marginpar{\colorbox{Author5}{\textcolor{white}{HP}} \textcolor{Author5}{#1}}}
\newcommand{\hemantIn}[1]{\colorbox{Author5}{\textcolor{white}{HP}} \textcolor{Author5}{#1}}

%Issue-wise commands	
\fi

%%
%% The "title" command has an optional parameter,
%% allowing the author to define a "short title" to be used in page headers.
%\title[]{Body Doubling in Virtual Reality: Understanding the Impact on ADHD Productivity, and Design Considerations}

\title[]{You Are Not Alone: Designing Body Doubling for ADHD in Virtual Reality}

\author{Zinat Ara, Imtiaz Bin Rahim, Puqi Zhou, Liuchuan Yu, Behzad Esmaeili, Lap-Fai Yu, Sungsoo Ray Hong}
% \affiliation{%
%   \institution{\textsuperscript{\textdagger} George Mason University, Fairfax, VA \country{USA}\\ \textsuperscript{\S}Temple University, Philadelphia, PA \country{USA}\\ \textsuperscript{\textdaggerdbl} The Ohio State University, Columbus, OH \country{USA}}
%   %\state{VA}
%   % \country{USA}
%   %\postcode{22031}
%   }

% \renewcommand{\shortauthors}{Redacted for blind reviews, et al.}
\renewcommand{\shortauthors}{Ara, et al.}

% %%
% %% The abstract is a short summary of the work to be presented in the
% %% article.

\begin{abstract}

Adults with Attention-Deficit/Hyperactivity Disorder (ADHD) experience challenges sustaining attention in the workplace. 
Body doubling, the concept of working alongside another person, has been proposed as a productivity aid for ADHD and other neurodivergent populations (NDs). 
However, prior work found no conclusive effectiveness and noted NDs’ discomfort with social presence. 
This work investigates body doubling as an ADHD-centered productivity strategy in construction tasks. 
In Study 1, we explored challenges ADHD workers face in construction and identified design insights. 
In Study 2, we implemented a virtual reality bricklaying task under three conditions: (C1) alone, (C2) with a human body double, and (C3) with an AI body double. 
Results from 12 participants show they finished tasks faster and perceived greater accuracy and sustained attention in C2 and C3 compared to C1. 
While body doubling was clearly preferred, opinions diverged between conditions. 
Our findings verify its effect and offer design implications for future interventions.

\end{abstract}

\maketitle
\section{Introduction}

% ADHD unemployment and challenges
A core characteristic of Attention Deficit/Hyperactivity Disorder (ADHD) is difficulty sustaining attention, which often undermines workplace performance and contributes to high unemployment rates in this population~\cite{campbell2023adhd, CRISTOFORI2019197, doyle2020neurodiversity}.
More than 15 million people in the US, approximately 1 in 16 adults, meet the diagnostic criteria of ADHD, with many more experiencing significant but subclinical levels of symptoms~\cite{staley2024attention, nh}.
Study results show that 60\% of this population is more likely to be unemployed from their job, 3 times more likely to quit impulsively, and 30\% more likely to face chronic employment difficulties~\cite{adda, doshi2012economic, hilton2009association}.
Attentional instability further makes them vulnerable when required to manage multiple demands, such as task switching, decomposing larger goals into subtasks, prioritizing actions, and shifting focus as circumstances change~\cite{chadd, CRISTOFORI2019197}.
They also face challenges completing routine tasks on time and regulating impulsive behaviors, which together can reduce productivity and make it harder to meet employer expectations~\cite{nadeau2005career, biederman2006effects, wheeler2018navigating}.

%what is body double
\textit{Body Doubling}, the practice of leveraging another person’s social presence to increase focus, motivation, and sustained engagement in tasks~\cite{sanders2024editor}, has emerged as a promising strategy to help individuals with ADHD extend their attention span~\cite{canela2017skills,al2020coping}.
In recent years, body doubling has gained recognition within neurodivergent (ND) communities as a potential productivity aid and is widely discussed across ADHD-specific platforms, social media, and online forums~\cite{adda, annavarapu2024comparative}.
Typically, a body double is a supportive individual who engages in a similar task or activity alongside the person they are assisting~\cite{annavarapu2024comparative}.
The underlying idea is that the neutral and calm social presence of another person can provide passive support to sustain engagement, reduce feelings of isolation~\cite{eagle2024something}, foster accountability~\cite{sanders2024editor}, and serve as a visual reminder to remain on task~\cite{eagle2023proposing}.

%% BIG BUT
While recent dialogue about body doubling suggests the potential of body doubling to improve attention span in tasks~\cite{eagle2024something,eagle2023proposing}, in applying the body doubling in practice, there exist two noteworthy challenges:
First, conventional body doubling relies on the physical presence of a peer, which can be logistically challenging and may provoke social anxiety, fear of judgment, and discomfort~\cite{bd,focus,campbell2023adhd,nadeau2005career}.
While being observed by another person can create accountability~\cite{eagle2024something} and enhance attentional focus~\cite{eagle2023proposing}, for someone anxious about being judged, the expectation to perform or avoid mistakes in front of a human observer can raise anxiety, even leading to task avoidance~\cite{edel2010alexithymia}.
In HCI, AI companions have been shown to reduce social anxiety in participants by offering non-judgmental, emotionally safe interaction with personalized support~\cite{oh2021systematic,nepal2024contextual} .
Second, no studies to date have clearly demonstrated the effectiveness of body doubling, and existing findings remain inconclusive~\cite{deshmukh2025toward,eugenia2024leveraging}.  
No peer-reviewed controlled trials have examined its impact on adults with ADHD, and the few available studies report limited outcomes.  
One study with ADHD participants found no significant effect~\cite{born2024effects}, while another with neurotypical (NT) individuals reported no behavioral differences across conditions~\cite{annavarapu2024comparative}.  
These signals highlight that key design factors for body doubling interventions remain unexplored and open.

In this work, we present two studies: Study 1 (S1) explores design insights for developing an effective body doubling intervention for adults with ADHD, and Study 2 (S2) evaluates its effects through a controlled experiment.  
We focus on construction work as the application area, since ADHD is the most prevalent ND condition among construction workers~\cite{adhdcons1,adhdcons}.  
We also employ a Virtual Reality (VR) environment, which reduces the discomfort of another person’s physical presence in body doubling scenarios and supports controlled experimentation, following recent HCI-oriented intervention studies with neurodivergent populations~\cite{adiani2022career,li2025generative}.  
Building on this rationale, we further examine how ADHD participants perceive differently when working with humans versus AI body doubles, an area not yet addressed in prior research.
Our overarching Research Questions (RQs) are as follows:
\begin{itemize}
    \item \label{rq1} \textbf{RQ1.} How body doubling concept can be designed as an effective strategy to support attention and productivity for individuals with ADHD in workplace contexts?
    \item \label{rq2} \textbf{RQ2.} To what extent can an AI-based body double elicit similar perceptions and outcomes as a human body double, and is it regarded as effective as its human counterpart?
\end{itemize}

%%% WHAT WE DID AND WHAT WE FOUND S1
S1 sought insights on how to ground body doubling design in the realities of construction work, particularly for adults with ADHD.
In doing so, we conducted a formative study with four construction personnel, including safety managers and workers with ADHD, and one research expert on ADHD.
S1 revealed 4 themes describing effective productivity approaches among workers with ADHD, (a) \textit{Companionship through modeling}, (b) \textit{Situational awareness}, (c) \textit{Repetitive task with clear guidance}, (d) \textit{Motivation through playful feedback}.
These insights provide 5 design considerations for developing body doubling interventions tailored to ADHD construction workers and directly informed our subsequent controlled evaluation in Study 2 (S2).

%%% WHAT WE DID AND WHAT WE FOUND S2
Based on S1 design insights, we employed 3 conditions for performing a virtual bricklaying task:
(C1) \textbf{Alone}; a baseline where participants complete the task without a body double;
(C2) \textbf{Human Body Double}; a condition where participants briefly interact physically with a human body double before entering the virtual environment and complete the task with him; and
(C3) \textbf{AI Body Double}; a condition where participants are told they will collaborate with an AI-driven, non-human agent (Wizard of Oz method) and complete the task with him.
12 participants with ADHD were recruited for the summative study (S2) to examine the effects of different body doubling conditions. 
S2 behavioral results showed that participants completed tasks faster in the presence of body doubles, while perceptual measures indicated that they reported greater accuracy and improved sustained attention under body doubling conditions. 
These findings validate the impact of both human and AI body doubles on ADHD productivity and point toward design implications for future interventions.
This work offers the following contributions:
\begin{itemize}
    \item \textbf{S1 Empirical Contribution:} Design insights that inform effective body doubling interventions for adults with ADHD in construction work.  
    \item \textbf{S2 Design Contribution:} Implementation of body doubling conditions, grounded in Study~1 insights, that can serve as a case instantiation for future body doubling designs.  
    \item \textbf{S2 Empirical Contribution:} Findings from a controlled study that explain how body doubling versus non-body doubling, and AI body double versus human body double, differ in behavior and perception.  
    \item \textbf{Implications for Design:} Potential directions for future research on developing body doubling as a broader intervention applicable within and beyond construction settings.  
\end{itemize}

\section{Related Work}

Research on ADHD productivity has largely emphasized individual strategies, such as planning tools, therapies, or self-regulation aids, with little attention to social approaches. 
More recently, body doubling—working alongside another person—has drawn attention as a potential productivity aid for ADHD.
Its effectiveness, however, remains elusive, with reports ranging from helpful to distracting.
Explaining this mixed signal may require understanding how ADHD traits shape attention and focus in the presence of others.
In what follows, we review (1) ADHD and productivity interventions, (2) body doubling and its unclear signals, and (3) ADHD traits that may explain why body doubling works for some but not others.

\paragraph{\textbf{ADHD and Productivity Interventions, a Landscape Dominated by Self-Directed Supports}}
Research on ADHD spans health, HCI, and social science, producing a rich body of interventions aimed at improving their productivity.
Yet, these efforts largely emphasize self-directed, individual-level support, with limited attention to socially grounded strategies.
Much of the prior work~\cite{asiry2018extending,rajarajeswari2024attention,lee2017adhd,phalke2023identification,avila2018towards} has investigated ways to sustain attention, regulate emotions, and organize tasks, key executive functioning challenges that directly shape work performance for ADHD individuals. 
Existing approaches include smartphone reminders, wearables, EEG devices, and AR/VR-based interventions that target attention retention and task management~\cite{spiel2022adhd,rajarajeswari2024attention,cuber2024examining}. 
For instance, a study was designed to assess how variations in text color (i.e., highlighting, contrast, and sharpening) affect the reading attention span of children with ADHD~\cite{asiry2018extending}.
Another study evaluated the effectiveness of a VR-based attention assessment tool that measured attention profiles by asking users to locate a target shape amid various distractions~\cite{rajarajeswari2024attention}. 
Cuber et al. investigated the potential of VR to increase attention span in a study with 27 university students with ADHD, who were asked to complete homework in a quiet virtual reading environment~\cite{cuber2024examining}. 
The results showed significant improvements in both concentration and motivation.
Similarly, O’Connell et al. designed a socially assistive robot that monitors attention and provides nonverbal feedback to college students with ADHD, which is perceived as helpful~\cite{o2024design}. 
Few studies explored the use of Augmented Reality (AR)
in improving the frustration tolerance of ADHD learners~\cite{ocay2018utilizing} while others have investigated the video-watching experiences of people with
ADHD to understand their video-watching frustrations and current strategies for access~\cite{jiang2025shifting}.
While these designs provide valuable support, they remain rooted in individual coping mechanisms. 
Broader productivity research also emphasizes tools and frameworks for general workforces~\cite{kim2019understanding,fritz2023cultivating,qian2024take,kobiella2024if}, yet these often overlook the behavioral and social needs of ADHD population. 
The importance of social contextualization~\cite{mikami2015importance}, such as peer support, team dynamics, accountability, and companionship, has been widely recognized as critical for productivity~\cite{nadeau2005career,mcintosh2023thriving,campbell2023adhd,qian2024take,kobiella2024if}. 
However, existing productivity interventions rarely incorporate or leverage these forms of social-contextual support.

\paragraph{\textbf{Body Doubling, Unclear Signals}}\
Body doubling, a productivity strategy that leverages another’s presence to support attention and task completion, has been proposed as a potential breakthrough for ADHD. 
However, controlled studies have yet to show conclusive effects, and its mechanisms of support remain unclear.
Despite its growing popularity in ADHD communities and commercial platforms such as \textit{Focusmate}, \textit{Flow Club}, and \textit{Flown}~\cite{adda,focus,flowclub,flown,dubbii}, empirical evidence is mixed.
In practice, people often engage in body doubling informally without labeling it as such, for example, working alongside a friend as an ``activity buddy'' or collaborating in professional settings through practices like pair programming, study buddy, etc~\cite{annavarapu2024comparative,lee2021personalizing,ren2025virtual}.
To investigate this, Eagle et al. surveyed 220 participants to understand how, when, and why people engage in body doubling~\cite{eagle2024something}.
They proposed the concept as a continuum of space/time and mutuality, where the presence of another person, whether in the same room, online, or even via media, serves as a form of accountability and reminder to stay on task~\cite{eagle2024something,eagle2023proposing}. 
Their findings show that body doubling helps people initiate, continue, or complete tasks by offering companionship, reducing feelings of overwhelm or anxiety, and leveraging subtle peer or social pressure~\cite{eagle2024something}.
Many individuals report that the presence of another person, whether a co-located peer, a remote partner, or even through “Study With Me” videos, helps them initiate tasks and sustain momentum by creating subtle accountability and companionship~\cite{eagle2024something,lee2021personalizing,studywithme}. 
Yet, experimental studies testing body doubles in controlled conditions show no significant effect of body doubling support, even if participants perceived them as motivating~\cite{annavarapu2024comparative,born2024effects}. Furthermore, existing body doubling tools often suffer from rigid scheduling, lack of interaction, or limited customization, which reduces their accessibility for people with ADHD~\cite{eugenia2024leveraging}. 
These limitations suggest that while body doubling is promising, more research is needed to clarify how and why it works, and under what conditions it can be most effective.

\paragraph{\textbf{ADHD Traits and Social Presence: Help or Hindrance}}
Difficulties with emotion regulation and sustaining attention often shape how individuals with ADHD perceive working alongside others.
This trait could lead to reluctance or even fear in social settings.
%% explaining about adhd traits social anxiety - this may explain why body dobule may not work
While group settings are known to motivate productivity in many populations, these social dynamics may not translate directly to ADHD.
Prior work shows that individuals with ADHD frequently experience stigma, fear of judgment, and heightened anxiety in social contexts~\cite{mcintosh2023thriving,campbell2023adhd}, which can undermine the potential benefits of co-working. 
For example, being randomly paired with strangers through online body doubling apps may feel uncomfortable or even overwhelming for some~\cite{eugenia2024leveraging}. 
Conversely, studies on virtual co-presence suggest that mediated environments, such as remote doubles or AI-driven avatars, can reduce social threat while still offering accountability and companionship~\cite{deshmukh2025toward,collins2024exploring}.
Unlike in-person body doubling, where social anxiety and fear of judgment~\cite{ara2024collaborative} may undermine productivity for some individuals, mediated doubles reduce the immediacy of social threat while still sustaining a sense of shared presence.
Careful calibration of social presence is also necessary, as excessive presence may induce stress~\cite{edel2010alexithymia}, whereas insufficient presence may fail to elicit motivation.
For instance, one study shows that co-working with avatars allows participants to feel “with others” without the discomfort of direct observation~\cite{collins2024exploring}, while others ~\cite{deshmukh2025toward,eugenia2024leveraging} explored how AI-driven companions can provide nonjudgmental presence and light-touch encouragement.
Despite promising conceptual frameworks, no peer-reviewed controlled research has examined whether AI doubles can replicate the motivational effects of humans while avoiding the social costs. 

\vspace{2mm}
In short, body doubling offers a promising alternative to the individual-focused tools that dominate ADHD productivity support.
Yet, as our review suggests, individuals with ADHD may also experience pressure or fear of being judged when working alongside others.
To mitigate these challenges, we identify two directions: (1) using virtual environments and (2) reducing social pressure by framing the body double as an AI rather than a human partner.
Beyond these, further design insights are needed to tailor body doubling interventions to specific contexts—such as construction work, where many adults with ADHD are employed.

\section{Positionality and Ethical Considerations}

We acknowledge the positionality of the research team. None of the authors identify as ADHD, though several have prior experience working with neurodivergent populations in educational, coaching, and research contexts. We recognize that this positionality shapes study design, data interpretation, and theme development. To mitigate bias, we drew on prior ADHD research, insights from formative studies with ADHD workers, and iterative feedback from neurodivergent advisors. All studies were conducted in the United States with adult participants residing there during the study period.  

We also recognize that research with ADHD participants in VR construction tasks presents risks at multiple levels, and we implemented protocols to protect participants.  

The first type of risk concerned collecting sensitive data on productivity, attention, and workplace challenges:  
(1) \textbf{Privacy and confidentiality}: All identifiable information was anonymized prior to analysis. Data were stored securely with encryption and access controls.  
(2) \textbf{Ethical oversight}: All study procedures were reviewed and approved by our university’s IRB.  
(3) \textbf{Informed consent}: Participants received IRB-approved consent forms in clear language. Consent was obtained prior to participation, and participants could withdraw at any time without penalty.  

The second type of risk related to participation in VR, including motion sickness, fatigue, or distraction:  
(4) \textbf{Physical comfort and safety}: Session lengths were limited, and participants were encouraged to take breaks or stop if discomfort arose.  
(5) \textbf{Psychological comfort}: We emphasized that there were no ``right'' or ``wrong'' outcomes, and mistakes were framed as part of the experiment.  
(6) \textbf{Clear instructions and support}: Structured guidance, visual cues, and reminders were provided to reduce confusion or overload.  
(7) \textbf{Accommodations}: Accessibility needs (e.g., breaks, VR sensitivity) were collected in advance, and adjustments were made accordingly.  

\section{Study 1: Formative Study}
S1 aimed to identify design rationales that could inform effective body doubling interventions and define an evaluation task suitable for the construction domain.
We focused on construction because ADHD is notably prevalent among construction workers~\cite{adhdcons1,adhdcons}.
\subsection{Method}
To gather insights, we conducted open-ended, semi-structured interviews with five participants who brought complementary expertise: construction safety managers, workers with ADHD, and an ADHD research expert.
These participants were well-positioned to inform us about both (1) the nature of construction work—characterized by hazardous, high-risk environments—and (2) the challenges faced by ADHD users, such as difficulties in sustaining situational awareness and attention.
Since our platform choice was Virtual Reality (VR), we also sought their perspectives on the opportunities and limitations of VR as a medium for implementing body doubling.

%Immersive technologies are already widely adopted in construction for safety training and for evaluating risk behaviors and team decision-making~\cite{pooladvand2025simulating}. 

\vspace{2mm}
\paragraph{\textbf{Participants}}
To capture multi-faceted perspectives on ADHD-related workplace challenges in construction, we recruited five participants. 
Two neurotypical (NT) safety managers who were experienced in supervising workers with ADHD, two construction workers who were clinically diagnosed with ADHD, and one was an expert NT researcher specializing in ADHD. 
Recruitment was conducted using a combination of convenience and snowball sampling. 
One of the authors drew on prior experience working with construction personnel to share contact information for potential participants.
Additional participants were referred through acquaintances with relevant experience.
In contacting them, we sent emails explaining the purpose of the study, the required experience to be a participant, compensation information of a \$25 gift card, and our contact point.
We also distributed a recruitment post in online groups and pages dedicated to ADHD construction workers.
Participants' ages ranged from 24 to 65 years.
Table 1 summarizes the details of our participants.

\renewcommand{\arraystretch}{1.2}%
\begin{table*}
\small
  \centering
  \begin{tabularx}{\textwidth}{p{0.04\textwidth} p{0.25\textwidth} p{0.04\textwidth} X}
  % \begin{tabularx}{\textwidth}{l >{\raggedright\arraybackslash}p{1.6cm} l X}
    \toprule
        \textbf{PID}
        & \textbf{Types}
        & \textbf{Age}
        & \textbf{Profile (years of experience)}\\
    \midrule
    %J
    M1 & Safety Manager (NT) & 52 & Corporate Safety Director, managing risk from safety standpoints (29 years)\\
    M2 & Safety Manager (NT) & 53 & A Senior Safety Manager who develops safety training and process (25 years)\\
    W1 & Construction worker (ADHD) & 65 & An electrician tech in building maintenance (42 years)\\
    W2 & Construction worker (ADHD) & 24 & A helper in construction plumbing service (1 year)\\
    E1 & Expert researcher (NT)  & 41 & Associate Professor and Licensed Psychologist specialized in ADHD (10 years)\\
  \bottomrule
  \end{tabularx}
  \vspace{2mm}
  \caption{
    A list of participants, from the left: (1) Participant ID, (2) Stakeholder type, (3) Age, and (4) their Profile and years of experience}
    \Description{A list of Study 1 participants that explains four types of information, their Participant ID, Stakeholder type, age, and their profile, along with years of experience}
  \label{tab:table1}
\end{table*}

\vspace{2mm}
\paragraph{\textbf{Interview and Analysis}}
All interviews were conducted online via Zoom by the first and the last authors.
At the beginning of each interview, we provided a consent form and asked participants to provide their agreement if they would like to proceed with the study. 
We recorded the conversation upon participants' agreement.
To ensure consistency across interviews, we used a slide deck in which questions were customized for each stakeholder role.

We structured the interview questions for safety managers (M1, M2) and construction workers with ADHD (W1, W2) into the following topics: 
(1) general work processes and safety concerns on construction sites,
(2) ADHD workers' challenges and coping strategies,
(3) practices of collaboration and teamwork, 
(4) key factors influencing workers’ performance in the field. 
In addition, we presented two short video clips from open-source VR construction sites that simulated field tasks and safety training scenarios to elicit (5)  their feelings about VR in this context and reflect on its potential to support workers’ performance in real-world settings.
For the ADHD expert (E1), the interview questions centered on three topics: (1) challenges that workers with ADHD encounter in the workplace,
(2) coping strategies they find helpful,
(3) expert point-of-view of approaches to help ADHD individuals sustain their attention and task performance, and
(4) potential designs of body doubling techniques that could better assist them.
The interviews lasted 50.7 minutes on average (SD=9.01) while the longest one lasted 62 minutes and the shortest one 42 minutes.
Every interview was transcribed by English-proficient transcribers without using AI transcription tools.

After data collection, we applied an iterative qualitative coding process. 
Two coders independently coded text segments from each transcript, then compared their insights and consolidated common findings. 
This process led to a unified thematic structure that reflected the results of S1. 

% \rayIn{Add more detail in coding process}
% \input{subsections/04_S1Bresult}
\subsection{Result}
We present our findings reflecting three stakeholder perspectives: safety managers, construction workers with ADHD, and an ADHD expert.
The study aimed to understand ADHD-related workplace challenges and the factors influencing worker performance. 
Through qualitative analysis, we derived four themes that describe effective approaches for sustaining attention and enhancing productivity among workers with ADHD.
We further detail how these themes informed the development of five design considerations (DCs), which address the RQ\ref{rq1}.

\subsubsection{Companionship through Modeling}
Although social anxiety was a concern, participants emphasized the effectiveness of collaborative environments and the role of companionship in supporting task performance. 
Companionship emerged as an important strategy for ADHD workers, offering accountability and concrete learning opportunities. As E1 explained, ``\textit{working collaboratively can also be kind of fueling for them…someone else can tap me on the shoulder when I’m zoned out},'' highlighting how co-working structures keep attention anchored. Peer-guided, hands-on learning was especially valuable, with W2 noting, ``\textit{He also has ADHD and he won’t explain things much, but he’ll kind of guide me through things, and do the task a bunch of times in front of me. And if I have any questions, he answers them}.'' Observing others also provided practical models to follow, as W2 shared, ``\textit{Watching my mechanic doing his job, I think of him as my model and it helps me going…I always needed a different way to learn than other people}.'' M1 emphasized the importance of adapting instructions to ensure understanding: ``\textit{So I may have to change how I present things…take a little bit more time to make sure that they’re understanding and not getting distracted. I want them to watch me and learn}.'' .

\paragraph{\textbf{DC 1. Provide sense of accountability reducing social pressure}}
Co-working arrangements, such as body doubling, establish an immediate sense of accountability through the presence of another person, helping individuals with ADHD remain anchored to their tasks. 
A body double can serve as a visual reminder or provide peer monitoring, fostering a sense of accountability for ADHD individuals, which can be achieved through either a real person or an intelligent agent form, in remote settings.
Translating this arrangement into a virtual environment introduces greater flexibility, reducing the intensity of direct social interaction.
In VR, this presence can be represented through avatars, characters, or even ambient supportive agents such as NPCs (Non-Playable Characters), which can provide a sense of shared engagement.

\subsubsection{Situational Awareness}
For individuals with ADHD, switching attention presents a significant challenge, particularly in high-risk settings (i.e., construction sites) where situational awareness requires monitoring both primary tasks and surrounding hazards.
The ability to shift attention and re-engage with the primary task is essential for successful multitasking, yet workers with ADHD often struggled to balance attention across tasks and environmental cues. M1 explained, ``\textit{So the extreme nature of the moving parts of a project all the time, it's about awareness, which is a key factor in ensuring someone's safety},'' while E1 emphasized that, ``\textit{Core symptoms of ADHD include difficulties controlling attention, such as being easily distracted or getting hyperfocused. Hyperfocus can prevent multitasking or recognizing environmental cues for needs to shift}.'' These challenges resulted in misunderstandings during high-stakes tasks. W2 recalled, ``\textit{The biggest problem I remember having was, my mechanic would tell me to cut a specific measurement, and I would ask him…what the hell are you cutting over here…you pay attention stupid!},'' noting how lapses in attention led to errors. They further added, ``\textit{If I'm doing a bunch of things at the same time, I kind of stack these things on top of each other…sometimes I'll forget the second step and skip to the third step}.'' Such accounts illustrate how distraction, hyperfocus, and overloaded working memory undermine performance.
This reinforces the need for external supports that help ADHD workers sustain awareness and manage cognitive load in complex environments. 
\paragraph{\textbf{DC 2. Support for divided attention between primary task and secondary events}}
In VR, introducing moving objects around the player can effectively simulate unpredictable on-site events, such as passing equipment or falling materials.
To better support attention, additional design features, e.g., adding sound effects, highlighting the object, or accelerating object animations, can be incorporated to make these events more salient and noticeable.

\subsubsection{Repetitive Task with Clear Guidance}
Although ADHD is associated with distractibility, participants (W2, E1) mentioned that individuals with ADHD can perform well on repetitive tasks in environments with clear routines, as repetition reduces the burden of constant decision-making. 
W2 explained, ``\textit{I guess a repetitive job is more helpful for me, I'm repeating it to myself as I'm doing it. 
If I'm doing a bunch of things at the same time…sometimes I'll forget the second step and skip to the third},'' while M1 added, ``\textit{Construction can be a very good environment for someone with ADHD, because some tasks are repetitive in nature and allow them to function as opposed to needing to put multiple things together at the same time}.'' Clear instructions were also critical, as W2 noted, ``\textit{If they can get a thorough understanding of like ‘Where does the pipe start? Where does it end? What does the pipe venting’…that can be helpful}.'' Early preparation was similarly described as foundational for reducing ambiguity, with M2 remarking, ``\textit{The more that you could expose folks in advance, make them feel more comfortable once they got out in a construction project for sure}.'' Together, these accounts show how repetition, paired with structured guidance and preparation, supports ADHD workers in managing attention and minimizing errors.  

\paragraph{\textbf{DC 3. Design repetitive task providing structured visual cues}}
In designing a repetitive task, it is needed to consider that task is not purely monotonous however, supported by external visual hints to help ADHD workers remain attentive and minimize errors.
Visual cues act as external scaffolds that reduce the need for constant self-monitoring and provide immediate guidance~\cite{shayesteh2022enhanced,hu2024exploring,cuber2024examining}.
For instance, providing a visual reference model or pattern allows participants to build something step by step through repetitive actions, while still demanding attentional engagement. 
Such designs blend predictability with attentional challenge, creating a balance that supports both focus and performance.

\subsubsection{Motivation through Playful Feedback}  
Participants highlighted that enjoyable environments and timely feedback were essential for sustaining engagement. W2 explained, ``\textit{I think having fun on the job is very important…like talking casual, bantering back and forth, and just having a good relationship with the people that you work with},'' and further noted, ``\textit{I think, having competing or like somewhat of a competition is healthy},'' framing competition as a motivator rather than pressure. They shared that watching others perform a task helps them both learn and stay engaged in their own work. Alongside playfulness, participants emphasized the need for continuous feedback to prevent distraction and uncertainty. M2, reflecting on VR training, remarked, ``\textit{The feature that does for fall protection…when you did it wrong, it shows the animation hitting the ground. So it gives that instant feedback that helps}.'' These accounts show that playful interaction, light competition, and consistent feedback together create motivating conditions that help ADHD workers maintain focus and productivity. 

\paragraph{\textbf{DC 4. Enable shared monitoring of task status}}
 When workers can track not only their own progress but also that of others, it creates a comparative frame of reference that anchors attention and sustains motivation~\cite{kuntsi2011intraindividual}. 
 In VR, this can be facilitated by placing players in view of one another where they can see each other, or incorporating progress bars that display task status. 
These features function as a reminder for continued focus and effort.
For the body double, whether human or agentic, the visible progress provides a social comparison cue. 

\paragraph{\textbf{DC 5. Introduce motivational dynamics}}
We found that making the workplace enjoyable and motivating is essential for enhancing productivity. 
As W2 noted, ``\textit{Giving a ranking or putting a rating system where you can rate players’ performance just like in fighting games. If you get a certain rank, if you beat somebody within a certain amount of time you get points}''. 
Introducing mechanisms, i.e., visible rankings, positional comparisons with other workers, or reward systems for achieving goals can sustain engagement and encourage persistence. 
Similarly, displaying virtual captions that reflect a player’s status or provide motivational feedback represents another design consideration that could be integrated into virtual body doubles to further support performance.

\section{Designing Body Doubles in VR}

Based on the synthesized result from the literature and S1 design considerations, we implemented three conditions to investigate the body doubling method, 
C1) Alone, C2) Human Body Double, C3) AI Body Double.
Following subsections present about designing evalaution task and each experiemental condition in details.

\begin{figure*}[t]
\centering
\includegraphics[width=\textwidth]{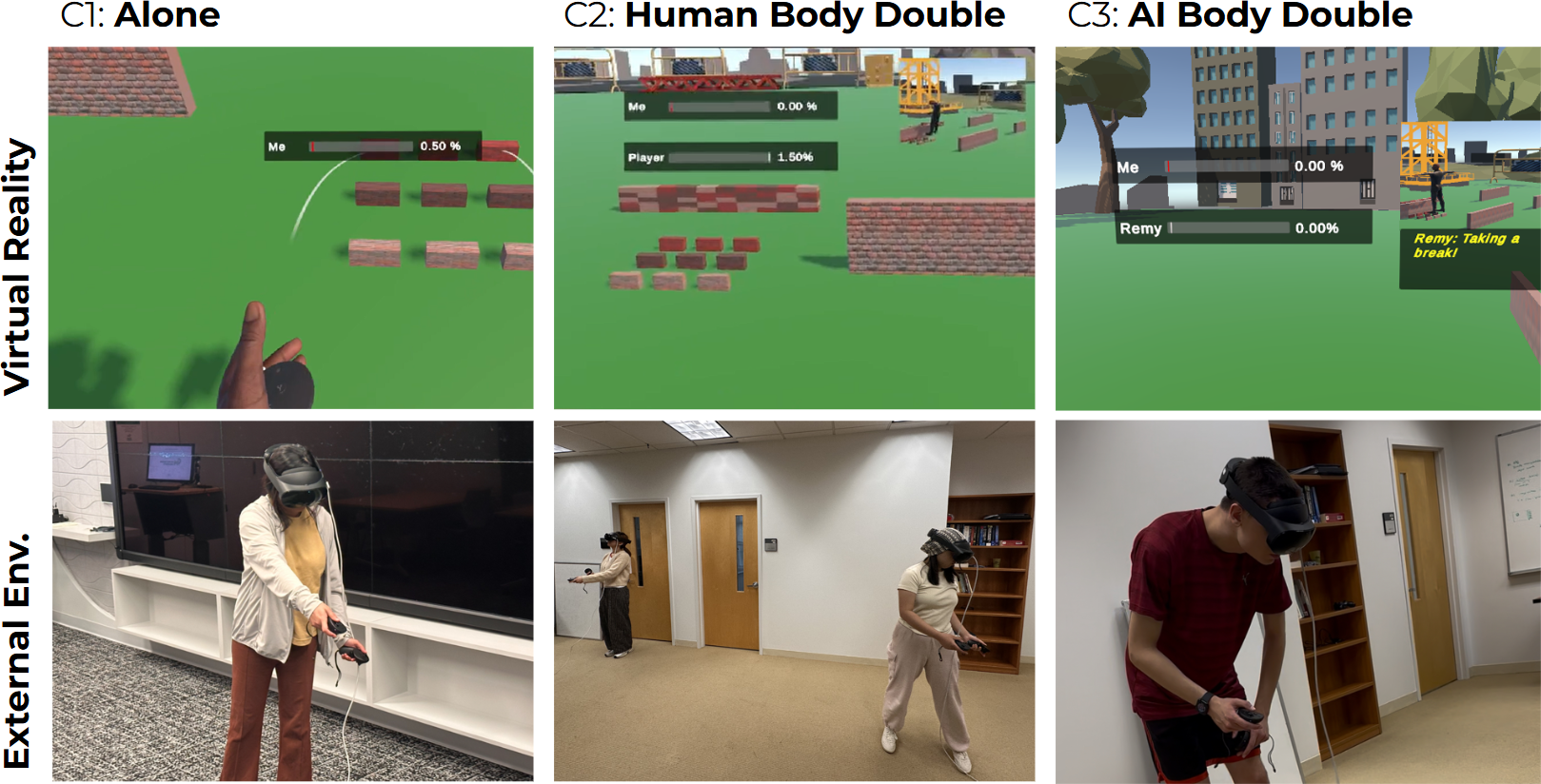}
\caption{\textbf{Experimental conditions of the VR bricklaying task:} (C1) Alone: participants completed the task without a body double; (C2) Human Body Double: participants worked alongside another person in the same VR environment; (C3) AI Body Double: participants worked with an AI-driven body double. The top row shows the VR interface with progress bars and secondary cam view, while the bottom row shows the corresponding external physical environment.}
\Description{A Figure that shows Experimental conditions of the VR bricklaying task: (C1) Alone – participants completed the task without a body double; (C2) Human Body Double – participants worked alongside another person in the same VR environment; (C3) AI Body Double – participants worked with an AI-driven body double. The top row shows the VR interface with progress bars and secondary cam view, while the bottom row shows the corresponding external physical environment.}
\label{fig:s2method}
\end{figure*}

\subsection{Task Description}
To explore body doubling effect, we developed a bricklaying task in VR derived from Study 1 design insights.
Participants have to perform this bricklaying task within a virtual construction site.
The virtual environment was designed to simulate a construction site with various construction equipment and objects.
The task involved constructing a wall by replicating a reference pattern composed of three different colored bricks. 
The pattern was generated by the random formation of 50 bricks of those three different colors.
Participants are presented with four empty wall bases and asked to place bricks one by one to match the given pattern. 
Each incorrect placement of a brick (wrong color or wrong position) was logged as a mistake.
When participants enter the VR scene, they can see the reference wall pattern, a pool of three types of colored bricks on the ground, and empty wall bases in front of them. 
To begin the task, the participant must pick the correct brick color from the pool and place it on the empty wall base using the VR hand controller. 
Each time a brick is taken, it is automatically replenished in the pool.
Once a brick is placed on the wall base, it cannot be retracted or removed.
Participants need to be careful not to grab the wrong brick. 
If an incorrect color is picked up, they can simply let it fall and then select another brick.
This process is continued until the wall is completed with 50 bricks that match the reference pattern. 
After finishing one wall, they can proceed to the next empty wall base and repeat the task.
Figure 1 shows the virtual environment and the task. 

There are three types of moving objects in the scene: a mixer truck, a cherry picker, and a crane. 
These objects move in a random manner and only one of them moves at a time. 
Each movement lasts between 20–30 seconds and then stops. 
After a short waiting period (any time between 45 - 80 seconds), another object (randomly selected by the system) starts moving again.
During the bricklaying task, participants also had to stay aware of their surroundings. 
Whenever they detected movement, they were required to give feedback by pressing a button on the controller.
Two of the objects (the truck and the cherry picker) included sound effects, while the crane movement was kept silent.

\subsection{Experimental Conditions}
We designed three experimental conditions to examine the effect of body doubling on task performance.
\begin{enumerate}[label=C\arabic*]
\item \textbf{Alone (baseline)}: Participants completed the bricklaying task without any body double present in the scene.
In the VR headset screen space (UI rendered relative to the player’s view rather than the 3D world), a progress bar is displayed to track task status, showing the participant’s progress as a percentage (e.g., 10\%). 
Figure 1.1 shows ..
\item \textbf{Human Body Double}: In this condition, participants briefly interacted with a human body double in the real environment before entering the virtual environment.
Participants were introduced to the human body double facilitator (another player) and informed that they would be playing with the body double in the next round.
Once both of them joined the scene in VR, the participant and the human double could see each other’s avatars, exchange a wave to say hello, and then begin performing the task side by side.
In the left corner of the VR headset screen space, two progress bars are displayed, one is for participant's self-progress another one shows body double's progress.
In the right corner of the participant's headset screen space, a secondary camera feed was displayed, allowing them to view the body double’s movements even when not directly looking at them. 

\item \textbf{AI Body Double}:
In this condition, participants were informed that they would be performing the task alongside an AI-driven body double, “Remy.” 
To ensure the avatar behaved consistently like a human player, we adopted a Wizard-of-Oz approach~\cite{1541964,bernsen1994wizard}, in which the avatar’s movements and actions were secretly controlled by a human operator. 
Similar to Condition C2, all supportive cues, including the secondary camera feed and progress bars were displayed in the VR headset screen.
To strengthen the perception of the body double as an autonomous agent, self-talking captions were incorporated. 
These captions were generated in response to the agent’s behavior, reflecting how it played and reacted in the environment and were designed to appear as if the agent was talking to itself rather than directly addressing the participant. 
This subtle self-commentary would create a sense of independent agency, while also motivating or reminding participants about their own tasks.
Three types of events triggered captions: (a) Idle mode: when the agent stopped working and stood idle, captions were shown randomly selecting from an idle pool, e.g., ``\textit{Calculating my next move…}'', ``\textit{Thinking…}'', ``\textit{Time to take some rest!}''; (b) Progress status: after crossing specific progress thresholds, captions were shown based on task advancement, e.g., ``\textit{That’s a good start!}'', ``\textit{Making progress…}'', ``\textit{Bricklaying is fun!}'', ``\textit{I gotta speed up…}'', ``\textit{We’re doing great!}''; and (c) Mistakes: when the body double made an error, captions were triggered from a mistake pool, e.g., ``\textit{Oops! made a mistake!}'', ``\textit{That didn’t go as planned}'', ``\textit{Incorrect placement :-( }''.

\end{enumerate}
Table 2 outlines the translation of formative insights from Study 1 into the design of Study 2, addressing our first research question (RQ1), which investigates how body doubling can be designed as an effective productivity strategy for workers with ADHD.

% \begin{table}[t]
% \centering
% \small
% \renewcommand{\arraystretch}{1.6}
% \begin{tabularx}{\linewidth}{%
%   % p{0.27\linewidth}%
%   % >{\raggedright\arraybackslash}X%
%    p{0.28\linewidth}%
%    p{0.32\linewidth}%
%   >{\raggedright\arraybackslash}X%
% }
% \toprule
% \textbf{Study 1: Subthemes} & \textbf{Study 1: Design Insights} & \textbf{Study 2: Features} \\
% \midrule
% \textit{Companionship, Visual Modeling, Procedural Adaptivity, Consistency} & Provide sense of accountability reducing social pressure & Virtual Human/Agentic body doubles, bricklaying in consistent rhythm \\
% \textit{Situational Awareness, Attention and Cognitive load} & Support for divided attention between primary task and secondary events & Moving objects in surrounding \\
% \textit{Good with Repetitive Tasks, Attention, Clear Guidance} & Design repetitive task providing structured visual cues & Virtual bricklaying task with pattern matching \\
% \textit{Progress Checking, Competitiveness} & Enable shared monitoring of task status & Progress bars + secondary camera feeds \\
% \textit{Self-regulation, Fun and Competitiveness} & Introduce motivational dynamics & Seeing body doubles'+ self progress, captions (Agent double) \\
% \bottomrule
% \end{tabularx}
% \vspace{2mm}
% \captionsetup{justification=raggedright,singlelinecheck=false}
% \caption{Mapping Study 1 Formative findings (subthemes) and Design insights to Study 2 System features}
% \end{table}
\begin{table}[t]
\centering
\small
\renewcommand{\arraystretch}{1.6}
\begin{tabularx}{\linewidth}{%
   p{0.5\linewidth}%
   >{\raggedright\arraybackslash}X%
}
\toprule
\textbf{Study 1: Design Considerations} & \textbf{Study 2: Features} \\
\midrule
DC 1. Provide sense of accountability reducing social pressure & Virtual Human/AI body doubles, bricklaying in consistent rhythm \\
DC 2. Support for divided attention between primary task and secondary events & Moving objects in surrounding, i.e, crane, truck, cherry picker \\
DC 3. Design repetitive task providing structured visual cues & Virtual bricklaying task with pattern matching \\
DC 4. Enable shared monitoring of task status & Progress bars + secondary camera feeds \\
DC 5. Introduce motivational dynamics & Seeing body doubles’ self-progress, captions (AI double) \\
\bottomrule
\end{tabularx}
\vspace{2mm}
\captionsetup{justification=raggedright,singlelinecheck=false}
\caption{Mapping Study 1 design considerations to Study 2 system features}
\Description{Table shows Mapping Study 1 design considerations to Study 2 system features}
\end{table}

\subsection{Implementation}
All system features were implemented in Unity 2022.3.38~\cite{unity}, using Unity’s built-in libraries and C\#.
Construction-related game objects were imported from the Unity Asset Store to create a realistic virtual construction site environment. 
The target VR device was the Meta Quest Pro~\cite{metaquest}. 
Meta’s SDK Interaction package~\cite{metasdk}, compatible with Unity, was utilized to build the core interactor and interactable components. 
These components supported the development of the bricklaying task and enabled interactions with the bricks, including grabbing, releasing, and placing them on the wall base. 
Meta building block components~\cite{metabb} were integrated to implement basic VR functionalities, including teleportation, camera rig setup, and player avatars (with body tracking) creation.
To enable multiplayer functionality, we used the Unity Netcode~\cite{netcode} library, which allowed body doubles to join the scene with the participant and perform the task in real time. 
For the AI body double condition, captions were generated using the GPT-5 model~\cite{gpt5} combined with rule-based logic to trigger context-specific events (e.g., idle, progress milestones, or mistakes), ensuring that the captions reflected the agent’s actions and maintained consistency across experimental conditions.

\section{Study 2: Summative study}
To understand the effect of body doubling on ADHD productivity, we conducted a summative study (S2) with the proposed designs.
Our S2 research questions aim to find as following:
\begin{itemize}
\item RQ1. Does the presence of a body double improve task performance and attention retention in individuals with ADHD, as opposed to only influencing perception of productivity?
\item RQ2. Can an AI body double be as effective as a human counterpart in improving task performance and attention retention for individuals with ADHD?
\item RQ3. What design factors should be considered for body doubles (either AI or Humans) to effectively improve task performance and/or quality of attention?
\end{itemize}
To address all these RQs, we adopted a mixed-methods approach that integrated both behavioral measures and perceptual data. 
We conducted an in-person laboratory study with 12 participants with ADHD, followed by post-task surveys and semi-structured interviews. 
The analysis of behavioral and perceptual outcomes from S2 further leads to design implications for future interventions.

\begin{figure*}[t]
  \centering
  \includegraphics[width=\textwidth]{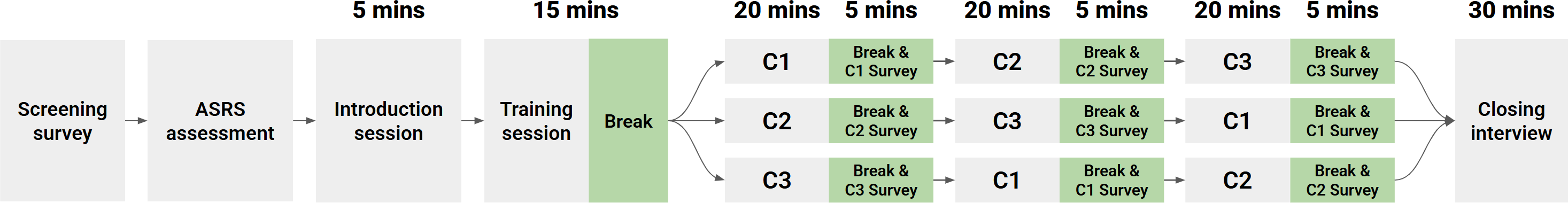}
  \caption{The flow of our summative study}
  \label{fig:s2method}
  \Description{A figure shows The Flow of our Summative Study}
\end{figure*}

\subsection{Method}

\subsubsection{\textbf{Participants}}
The study was advertised through a flyer approved by the Student Center directors and posted in various campus locations. 
Additionally, we asked acquaintances and colleagues to share the flyer with potential participants.
We recruited 12 participants between the ages of 18 and 31 with a mean age of 22.75 years. 
Out of 12 participants, 5 were identified as female and 7 as male. 
The majority of participants were current university students, some were employed in university staff positions, and the remainder were from other professions.
Participants were required to be at least 18 years old and physically attend the university where the study was conducted. 
Before recruitment, participants had to complete a screening process including a survey and ASRS assessment~\cite{Association,edition1980diagnostic} consists of 18 questions focused on two main symptom domains, inattention and hyperactivity-impulsivity.
Based on survey responses and assessment scores, we recruited 12 eligible participants, of whom 8 reported prior experience with VR.
None of the participants reported a history of discomfort or motion sickness.
Participants received \$100 in the form of a gift card for completing the experiemental study, post session surveys and follow-up interview which together lasted approximately two hours. 
The study was approved by the university Institutional Review Board (IRB) and all participants signed informed consent form.

\subsubsection{S\textbf{creening Survey and ASRS assessment}}
The flyer included a QR code that directed individuals to a screening survey. 
The survey collected demographic information and eligibility details, including prior ADHD diagnosis, experience with VR, and health conditions such as physical discomfort or motion sickness. 
These questions were used to identify potential medical concerns that could compromise participant safety and well-being during VR engagement.
Interested individuals were also asked to complete the ASRS v1.1 assessment~\cite{Association,edition1980diagnostic}, a diagnostic tool developed by the World Health Organization (WHO) in collaboration with researchers at Harvard Medical School for the evaluation of Attention-Deficit/Hyperactivity Disorder (ADHD) in adults. 
The ASRS v1.1 is widely used for both screening and monitoring the severity of ADHD symptoms over time, with all questions designed for self-report to capture individuals’ experiences directly.
The instrument includes 18 items aligned with DSM-IV-TR criteria~\cite{edition1980diagnostic}, divided into two symptom domains: inattention and hyperactivity–impulsivity.
Survey responses and assessment scores were reviewed by the researchers and 12 participants were selected for Study 2, who met the eligibility criteria.
We contacted eligible participants and provided them with the IRB-approved consent form, which describes study procedures and research protocols in detail. 
Participants who agreed to the consent form were given access to an online shared timetable where they could book time slots that fit to their schedules.
We scheduled in-person sessions by sending calendar invitations based on participants’ stated availability.

\subsubsection{\textbf{Pre-study setup}}
Prior to running the main study, we conducted several dry-run sessions to test the system and refine the study protocols. 
A key objective of these sessions was to establish consistent procedures for the human body double facilitator so that performance remained comparable across participants.
We conducted dry runs with two neurotypical student volunteers (one with prior VR experience, one without) and one student volunteer with ADHD (experienced with VR).
This process helped verify our recruitment criteria, ensuring alignment with the study sample, in which some participants had no prior VR experience.
Each volunteer was instructed to complete the bricklaying task by building four wall bases in the scene. 
Completion times were recorded and averaged to determine the pacing baseline for both the human and AI body doubles. 
This ensured that their behaviors and actions followed a consistent rhythm during the study. 
One volunteer later agreed to serve as the human body double. 
Through practice sessions, this volunteer rehearsed a standardized protocol, for example, counting from one to eight before picking up and placing a single brick, repeating this cycle throughout the session.

The same protocol was applied to the agentic body double. 
One of the authors performed this role (Wizard of Oz technique), following the same pacing rules as the human double. 
Additional scripted events were introduced for the agent to simulate varied behavior, such as idling for 10 seconds after reaching 10\% task progress or intentionally making an error after five minutes of play. 
These events were designed to trigger the corresponding captions. 
A researcher monitored the session and provided timed reminders to ensure that both body doubles followed the predefined protocols consistently.
\subsubsection{\textbf{Study Procedure}}
To conduct our study, we used Meta Quest Pro headsets connected to two computers. 
One machine was equipped with an Intel Core i9-11900K processor, a GeForce RTX 3080 Ti GPU, and 128 GB of RAM, while the other had an Intel Core i7-9700K processor, a GeForce RTX 2080 Ti GPU, and 32 GB of RAM.
At the start of each session, a researcher assisted participants in adjusting the VR headset for comfort and clarity. 
Participants were first introduced to a training scene that contained a simple patterned reference wall, several bricks, and a wall base. 
This scene familiarized them with the bricklaying task and the VR interaction mechanics, such as grabbing, releasing, and placing bricks using the controller. 
Once participants completed the training and demonstrated familiarity with the process, they proceeded to the main study sessions.

Each participant experienced all three conditions (C1, C2, and C3). 
The order of conditions was counterbalanced across participants using a Latin square design~\cite{bradley1958complete} to mitigate potential learning effects in the within-subjects setup. 
During each condition, participants were also instructed to remain aware of object movements in the environment and press a designated button on the right controller whenever movement was detected.
Before each session, participants were introduced to the upcoming condition. 
In the baseline condition (C1), they were told they would complete the task alone. 
Here, they could track only their own progress. 
In the human body double condition (C2), participants were introduced to the double in person prior to the session and engaged in a brief interaction. 
Once both players entered the VR environment, they could see the avatars and greeted each other with a wave before beginning the task. 
Each player worked independently on separate wall bases with their own brick sets, while sharing the same reference wall pattern. 
Participants could observe their own and the body double’s progress via progress bars and monitor the double’s activity through a secondary camera feed.
In the agentic body double condition (C3), participants were told they would perform the task alongside an AI-controlled player, “Remy.” 
Once both avatars entered the VR scene, the task began. 
Participants could again view dual progress bars and the secondary camera feed. 
In addition, Remy’s presence was augmented by self-talk captions that appeared during play, designed to reinforce the perception of autonomy and provide motivational or situational commentary.

Each session lasted 20 minutes, as prior studies indicate that VR sessions of 20–40 minutes balance engagement and comfort, while longer durations increase risks of fatigue, eye strain, and motion sickness~\cite{kourtesis2019validation,vrtime}. 
Five-minute breaks were provided after each session, during which participants completed the post-survey. 
Participants were also encouraged to take additional rest as needed until they felt fully comfortable before continuing.
They were also instructed to notify researchers immediately if any discomfort arose during a session.
The entire study lasted approximately two hours, including training, main study sessions, and post-study data collection.
\subsubsection{\textbf{Post-study}}
After completing each session, participants were asked to fill out a post-session survey. 
The survey contained five questions designed to capture user perspectives on task efficiency, task accuracy, quality of attention and focus, situational awareness, and task continuity for that session. 
All items were presented on a 5-point Likert scale, with response options ranging from 1 ( ``Strongly Disagree'' ) to 5 ( ``Strongly Agree'' ).

Following the completion of all sessions, we conducted a 30-minute semi-structured interview to further explore participants’ experiences. 
The interview probed their perceptions of differences across conditions in terms of efficiency, accuracy, and sustained attention.
Participants were also asked to compare their experience of completing the task alone versus with body doubles, to reflect on differences between the human and AI body doubles, and to provide feedback on the design of supportive cues such as progress bars, secondary camera views, and captions.
Finally, we asked participants about their preferences for future design features they would like to see in body doubling systems.

\subsubsection{\textbf{Data Collection and Analysis}}
From the VR sessions, we collected behavioral data related to task efficiency, task accuracy, and object detection performance. 
Specifically, we logged the number of bricks placed with timestamps, the total count of bricks placed, brick–pattern matching percentage, number of mistakes, timestamps of moving-object events, and participants’ detection feedback with timestamps. 
Participants' movement and eye gaze data were also recorded for each session.
Post-session surveys captured perceptual data on participants’ subjective experiences, including perceived task efficiency: speed of completion, task accuracy: quality of wall construction, sustained attention: level of focus, situational awareness: ability to identify moving objects, and task continuity: ease of resuming bricklaying after detecting object movement. 
Post-study interviews complemented these data by probing participants’ reasoning behind their ratings, their overall impressions of the body doubling conditions, comparative reflections across conditions, and feedback on design features.
For analysis, we defined three primary performance metrics: 
1) \textbf{Task Efficiency}: the speed of task completion, calculated as the number of correctly placed bricks per minute. 
Correctly placed bricks are defined as the total number of bricks a participant placed, minus the number of mistakes they made.
% % Efficiency formula
% \[
% \text{Efficiency (bricks per minute)} = 
% \frac{\text{Number of Correctly Placed Bricks}}{\text{Time in minutes}}
% \]
% \noindent
% Where:
% \[
% \text{Correctly Placed Bricks} = 
% (\text{Total Bricks Placed} - \text{Mistakes})\]
% \[
% \text{Time in minutes} = 
% (\text{Timestamp of the last brick placed} - \text{Timestamp of the first brick placed})
% \]
2) \textbf{Task Accuracy}: the precision of wall construction, measured as the proportion of correctly placed bricks out of the total number of bricks placed.
% Accuracy formula
% \[
% \text{Accuracy} = 
% \frac{\text{Correctly Placed Bricks}}{\text{Total Bricks Placed}} \times 100\%
% \]
3) \textbf{Object Detection Accuracy}: the timeliness and correctness of detecting moving objects, calculated as: 
% Score formula
\[
\text{Score} = \frac{100}{N} \left[(K - M) + \sum_{j=1}^{M} \max \bigl(0,\, 1 - \alpha(d_j - T)\bigr)\right]
\]

\noindent where $N$ = total events, $K$ = correctly detected events, $M$ = late detections (after threshold $T$), $d_j$ = delay in seconds for each late detection, $T$ = threshold (10s in our study, with no penalty within this time), and $\alpha$ = penalty rate per extra second of delay (set to 0.1). 
This scoring gives full credit for on-time detections, partial credit for late detections (scaled by delay), and no credit for missed events. 
$\alpha$ = 0.1 is set to emphasize the role of rapid detection in our task, which helps us see the differences clearly between different conditions.
From post post-study survey, we also analyzed \textbf{Sustained Attention} (perceived focus) and \textbf{Task Continuity} (perceived task resumption) metrics.

\subsection{Quantitative Results}

In this section, we report findings on participants’ performance behaviors on the task and its associated performance metrics. 
For each metric, we present quantitative results based on our analysis on data collected from the VR sessions and post-study surveys and contextualize these outcomes with the post-study interviews.

\subsubsection{Task Efficiency}
This metric captures how fast participants completed the bricklaying task. 
We first examined the data distribution to determine the appropriate statistical analysis. 
Task efficiency data collected from the VR sessions followed a normal distribution. 
Given the within-subjects design, where each participant experienced all three conditions (C1, C2, C3), we applied a parametric repeated-measures ANOVA.
The analysis revealed a significant effect of condition on efficiency, $F(2,22) = 6.51, p = 0.006$. 
Post-hoc comparisons with Bonferroni correction showed that efficiency was significantly higher in both the human double condition (C2, p = 0.040) and the AI double condition (C3, $p = 0.030$) compared to the baseline (C1). 
No significant difference was observed between C2 and C3 ($p = 1.000$).
These findings suggest that body doubles, whether human or AI, significantly improved task efficiency relative to working alone.
The mean efficiency in the baseline condition was 8.49 bricks per minute, which increased to 10.82 in the human double condition and 11.06 in the AI double condition. 
This reflects mean differences of 2.33 (C1–C2) and 2.56 (C1–C3), indicating that participants laid bricks faster and more consistently when accompanied by a body double. 
The difference between C2 and C3 was minimal (0.24), demonstrating that both forms of doubling provided nearly equivalent benefits. Effect sizes (Cohen’s $d_z$) reinforced these findings, the comparisons between baseline and human double ($d_z = -0.85$) and baseline and AI double ($d_z = -0.90$) were both large effects, while the human and AI comparison effect ($d_z = -0.09$) was negligible.

To analyze perceptual efficiency from the post-study survey question `` \textit{In this session, I was able to complete the task faster}'', we applied a non-parametric Friedman test, followed by Wilcoxon signed-rank tests for post-hoc comparisons.
The Friedman test for perceived efficiency did not reveal a statistically significant effect of condition, $\chi^2(2) = 3.38$, $p = 0.185$, indicating when comparing participants’ efficiency across the three conditions no significant effect is found.
On average, participants rated their efficiency as 3.00 when working alone, increasing to 3.50 with a human double and 3.75 with an AI double. 
The mean differences suggest that participants perceived themselves to be somewhat more efficient when accompanied by a body double, with the largest gain observed between baseline and the AI double condition (C3-C1 = 0.75).
Effect sizes (Cohen’s $d_z$) comparison between C1 and C2 ($d_z = -0.40$) represents a small-to-moderate effect, while the C1 versus C3 comparison ($d_z = -0.71$) suggests a medium-to-large effect, showing that participants perceived a better efficiency boost with the AI body double support.
The behavioral data from VR sessions showed clear and statistically significant gains in task efficiency. 
In contrast, the survey data revealed no significant differences across conditions, indicating that participants’ subjective perceptions did not always align with their actual performance. 
They may have underestimated or failed to notice differences in efficiency across conditions.
However, interview feedback told a different story, several participants reported that working with body doubles felt more effective than working alone, consistent with the behavioral results. 
A likely explanation is that the Likert-scale survey format lacked sensitivity, as many participants selected mid-range or similar scores across conditions, reducing the chance of detecting statistically significant effects.

\subsubsection{Task Accuracy}
This metric assessed how accurately participants replicated the reference wall pattern during the bricklaying task. 
As the behavior data were not normally distributed, we applied a non-parametric Friedman test with Wilcoxon signed-rank tests for post-hoc comparisons.
The result indicated that there was no significant effect of condition on accuracy, $\chi^2(2) = 3.17$, $p = 0.205$. 
This suggests that participants’ accuracy in placing bricks remained relatively stable across all conditions and post-hoc Wilcoxon signed-rank tests confirmed whether participants worked alone or with presence of a body double, their accuracy scores did not differ in a statistically meaningful way.
The Average accuracy was 86.57\% in the baseline condition, increasing to 90.81\% with a human double and 88.54\% with an AI double. 
The largest mean difference was observed between baseline and the human double condition (C2-C1 = 4.24).
Effect size comparison between baseline and human double yielded a small-to-moderate effect ($d_z = -0.40$), suggesting that participants were somewhat more accurate when supported by a human double. 
The baseline vs. AI double effect size was smaller ($d_z = -0.21$), indicating only a slight improvement compared to working alone.

While analyzing the perceived accuracy collected from the survey question `` \textit{In this session, I was able to finish the task more accurately}'', Friedman test revealed a significant effect of condition on accuracy, $\chi^2(2) = 9.39$, $p = 0.009$, indicating that participants’ accuracy scores varied meaningfully across the three conditions.
Post-hoc comparisons clarified that accuracy was significantly perceived higher both in the human ($p = 0.007$) and AI ($p = 0.045$) body double conditions compared to the baseline.
Participants’ survey and interview responses suggested that they felt body doubles improved the quality of their task performance, even though this perception was not strongly reflected in the behavioral data.
One possible reason could be, the presence of a human or AI partner may have made participants feel more attentive and engaged, leading them to perceive their performance as higher quality, regardless of whether their objective accuracy scores improved significantly.
The divergence between perceptual and behavioral outcomes suggests that body doubles influence not only ``\textit{how work is measured}'' but also ``\textit{how work feels}''.

\subsubsection{Object Detection Accuracy}
This metric evaluated participants’ divided attention, specifically their situational awareness and ability to detect object movements in the environment.
The Friedman test for behavioral object detection did not reveal a significant effect of condition, $\chi^2(2) = 0.17$, $p = 0.92$. 
This means that participants’ ability to correctly notice and identify moving objects was statistically similar across all the three conditions.
From the perceptual data collected for the survey question, ``\textit{In this session, I was able to notice and correctly detect when surrounding objects moved}'', we observed the same trend, $\chi^2(2) = 2.00$, $p = 0.368$ with no difference among the conditions.
Participants’ performance was relatively stable across conditions, with only small variations in the mean scores.
The mean differences indicate that participants performed marginally better with a human double compared to baseline (C2-C1 = 2.95) but slightly worse with the AI double compared to baseline (C3-C1 = -1.74).
The human versus AI double comparison showed a small effect size ($d_z = 0.26$), suggesting that participants may have detected objects a little more reliably when paired with a human double, though the difference was still weak.
This result contrasts with the findings for efficiency and accuracy, where body doubles produced significant improvements. 
One possible explanation is that object detection relies more on moment-to-moment attentional shifts rather than overall task pacing or precision, which can be influenced by the motivational boost of having a body double.
Vigilance in noticing unexpected movements is less likely to benefit from social or agentic support.
This insight is also aligned with the notions participants shared in the interviews, some reported that they could notice surrounding events better when working alone, as they were less absorbed in the collaborative aspect of the task, while others felt that with a body double present they were more engaged with the task and less attentive to their environment.
Another reason for detection was the sound effect feature, majority of the participants missed detect when the object ``Crane'' moved, which was silent. The truck and cherry picker made noises which was easy attracted their attention.

\subsubsection{Sustained Attention}

This metric assessed participants’ self-reported focus, indicating sustained attention and perceived quality of attention across all conditions, related to the survey question ``\textit{In this session, I was able to stay focused on the task}''.
The Friedman test for perceived focus revealed a significant main effect of condition, $\chi^2(2) = 18.82$, $p < 0.001$, indicating that participants’ self-reported focus varied reliably across the three conditions. 
This suggests that the presence of a body double had a meaningful influence on how attentive and engaged participants felt while completing the task compared to when they worked alone.
Post-hoc Wilcoxon tests with Bonferroni correction clarified participants reported significantly higher focus in both C2 and C3 compared to the baseline, with adjusted p-values well below 0.01. 
This finding shows that body doubles consistently enhanced participants’ ability to stay concentrated on the task, regardless of whether the double was human or AI. 
However, there was no significant difference between the human and AI doubles.
The result verifies that body doubling plays an important role in sustaining attention in task environments and both human and AI doubles helped participants feel more engaged and less distracted.
This aligns with participants’ qualitative remarks, that seeing body doubles doing their task with its progress, created a strong sense of accountability, which in turn motivated the participants to pay more attention to their own tasks.

\subsubsection{Perceived Task Continuity}

Friedman test for perceived task continuation, based on the survey item ``\textit{In this session, I was easily able to get back to bricklaying after noticing the object movements}'', showed a significant effect of condition,$\chi^2(2) = 18.82$, $p < 0.001$. 
This result indicates that participants’ perceptions of how easily they could resume the bricklaying task after distractions
Post-hoc Wilcoxon tests revealed that participants rated task continuation significantly higher in both the human and AI body double conditions compared to the baseline.
The scores suggest that body doubling enhanced participants’ perceived ability to maintain continuity in their work, even when external distractions were present.
Interviews confirmed that working alone made it more difficult and less motivating for participants to quickly re-engage with the bricklaying task after noticing object movements. 
In contrast, the presence of a body double introduced a sense of pressure and accountability, which reinforced participants’ motivation to return to the task.
This confirms the strong positive reinforcement of the body doubling effect on sustaining attention for individuals with ADHD.

\begin{figure*}[t]
  \centering
  \includegraphics[width=\textwidth]{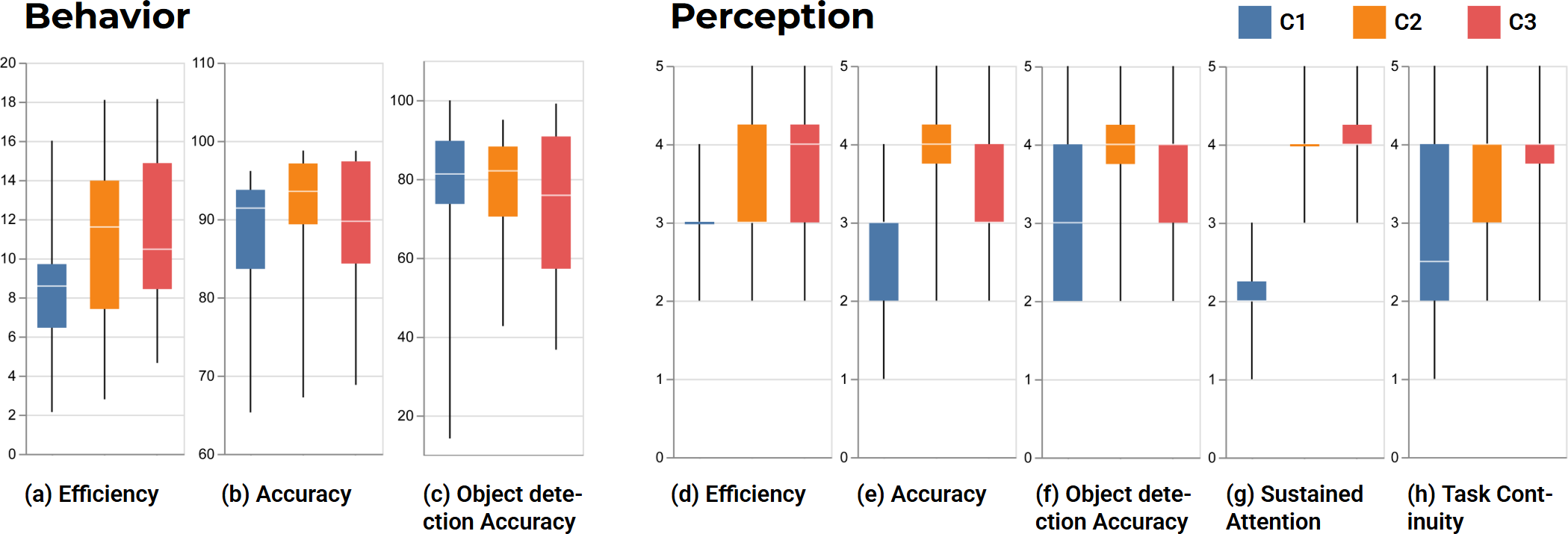}
  \caption{\textbf{Summative study result for each experimental conditions:} a) Behavioral Efficiency: box plots for no. of bricks laid per minutes, b) Behavioral Accuracy: box plots for task accuracy score in \%, c) Behavioral Object Detection Accuracy: box plots for object movement detection accuracy score in \%, d) Perceived Efficiency: box plots for users' perceptual efficiency ratings on a scale of 1 to 5, e) Perceived Accuracy: box plots for users' perceptual accuracy ratings on a scale of 1 to 5, f) Perceived Object Detection Accuracy: box plots for users' perceptual object detection accuracy ratings on a scale of 1 to 5, g) Sustained Attention: box plots for users' perceptual focus ratings on a scale of 1 to 5, h) Task Continuity: box plots for users' perceptual task resumption ratings on a scale of 1 to 5.}
  \label{fig:s2method}
  \Description{Summative study result for each experimental conditions: a) Behavioral Efficiency: box plots for no. of bricks laid per minutes, b) Behavioral Accuracy: box plots for task accuracy score in \%, c) Behavioral Object Detection Accuracy: box plots for object movement detection accuracy score in \%, d) Perceived Efficiency: box plots for users' perceptual efficiency ratings on a scale of 1 to 5, e) Perceived Accuracy: box plots for users' perceptual accuracy ratings on a scale of 1 to 5, f) Perceived Object Detection Accuracy: box plots for users' perceptual object detection accuracy ratings on a scale of 1 to 5, g) Sustained Attention: box plots for users' perceptual focus ratings on a scale of 1 to 5, h) Task Continuity: box plots for users' perceptual task resumption ratings on a scale of 1 to 5.}
\end{figure*}

\subsection{Qualitative Findings}

In this section, we present our findings from post-study interviews, where participants reflected on their study experiences and shared perspectives on the effectiveness of body doubles for ADHD productivity and preferences on body double types.

\subsubsection{Impact on Productivity: Alone vs Body Double}

The majority of the participants (10 out of 12) mentioned, working alone led them to be distracted, reduced their task speed, and sustained attention. 
Some participants reported feeling more relaxed; however, this state was also accompanied by more mistakes, as P1 reflected, ``\textit{Alone I was more relaxed, I was less focused, I made more mistakes as opposed to the other two settings}''.
P3 mentioned, ``\textit{When I was working alone 100\% far easier to get distracted by things moving around and noises especially}''.
This sense of working in isolation also contributed to slower progress as P2 said, ``\textit{Alone, I sometimes felt like I was in my own world . I was able to get things done my own pace, which takes longer than usual}''.
Absence of external accountability can reduce motivation leading to task procrastination.
P6 explains that working alone created a ``lack of pressure'' to maintain pace, which made it easier to slow down or disengage from the task.
By contrast, the presence of a body double, whether human or AI, was repeatedly linked to greater focus, energy, and motivation. 
Participants reported that the body double acted as an accountability mechanism, reminding them to stay on task and reducing tendencies to drift or lose time.
P8 explained, ``\textit{Definitely having a body double, it did help me have more concentration on tasks, it did help me stay more focused, be more energetic}''.
Similarly, P7 noted, ``\textit{Alone, I can get easily off the task, but with a body double I’m reminded again that all I have to do is bricklaying and someone is there observing me}''.
The sense of being observed created a productive pressure that kept them moving forward.
One participant (P4) believed they were more focused when working alone, noting that the presence of another person was distracting. 
Another participant (P10) reported no significant difference in productivity, however, described the task as more enjoyable when performed with a body double.
This suggests how the impact of body doubling on productivity is shaped by individual preferences, attentional styles, and comfort with social presence.

\subsubsection{Body Doubling Preference: Human vs AI}

We observed divergent participant perspectives regarding their preferences for AI versus human body doubles.
This difference stemmed from two contrasting perspectives: the ``realistic social pressure'' associated with human body doubles and the sense of ``freedom from judgment'' provided by AI body doubles.
For some, human presence was a stronger motivator, creating accountability and competitiveness that directly enhanced productivity. 
P12 mentioned, ``\textit{Definitely, I felt fastest with the human body double..I felt a bit more of a presence there}''.
P2 expressed greater sustained attention with the human influence, `\textit{`Human body double did a lot more to become more focused as opposed to the AI body double which seemed to be a bit boring to be around}''.
At the same time, other participants found human doubles introduced too much pressure, making them feel awkward or distracted. 
For these individuals, AI was the preferred option because it allowed them to work without the anxiety of being judged. 
As P4 shared, ``\textit{I felt a little awkward with the human presence but with AI it was a lot less pressure, I didn’t have to worry about the stress of being watched}''. 
P8 liked the AI as a comforting companion, remarking, ``\textit{It’s just knowing that it’s an AI. It’s like the equivalent of having a stuffed animal to talk to}''.
These reflections reveal that AI doubles can create a form of accountability that supports focus while minimizing social discomfort.
Participants also differentiated between task types in their preferences. 
For repetitive or “mindless” work, AI was perceived as sufficient to keep the task moving, as p1 said, ``\textit{When you’re doing something kind of mindless or repetitive, then it would be nice to have an AI body double to just make sure you stay on track}''. 
However, in tasks requiring deeper concentration, such as studying, some participants felt that human doubles were more effective, as P3 mentioned, ``\textit{For studying I think I would actually prefer a human. Now I have a pressure to kind of fit in and study, because AI doesn’t care}''. 
This suggests that preferences are context-sensitive.
Participants reflected on how body doubling mirrors everyday social dynamics of productivity. 
P9 noted, ``\textit{I do think having a person with me does help me get certain tasks done..if you’re in a room full of people doing the same thing, I think it’s a lot easier to do whatever you’re doing rather than just being by yourself completely}''. 
Yet others admitted they could easily ``ignore'' the AI because it felt less real, P11 said,``\textit{You can kind of easily ignore it..because it’s just a computer}''.
Some participants also reported that the AI body double supported their learning of the task. 
As P5 explained, ``\textit{I watched him doing, making walls, laying bricks using columns, I thought that was a good idea to start}''.

\subsubsection{Feedback on Supportive Cues}

The progress bars were consistently considered as a useful tool for maintaining awareness of personal and partner performance.
Participants valued being able to track their own pace as well as compare it to their body double’s progress. 
P2 described, ``\textit{Knowing the progress bar would help you move forward… you can change your pace according to that value}''. 
Others emphasized the motivational effect of comparison: P9 said, ``\textit{Seeing a progress bar..it does also help productivity. It’s like, oh, I’m doing better compared to the others}''. 
This competitiveness created positive pressure for many: as P4 said,``\textit{The competitiveness, the pressure, these kind of things actually helps. It definitely makes you go faster}''. 
However, participants also recognized potential downsides, acknowledging that comparisons could discourage some people if they felt behind.
Only few participants perceived the secondary camera view was generally positive, as P3 suggested, ``\textit{The secondary cam was kind of nice.. it felt more like a background simulation which helped me to keep going}''.
P8 appreciated the convenience of being able to glance at their partner without disrupting their own focus, ``\textit{I thought cam was helpful because I didn’t have to go all the way and be like, where are you doing? You can kind of see what they’re doing}''.
However, the majority of participants (8 out of 12) reported paying little attention to the feature, preferring instead to observe the body doubles directly. 
As P7 noted, ``\textit{Honestly, I didn’t look at it that much, because I could see them directly if I wanted to watch}''.
Participants viewed AI captions as encouraging and socially supportive, though reactions were mixed depending on content. 
Some appreciated captions that framed the task positively, noting that they made the activity feel less monotonous: P8 mentioned, ``\textit{They were talking about how this task is not a boring task at all—they’re trying to make it less boring}''. 
Others described captions as offering companionship, P4 said, ``\textit{I noticed the little chats from the AI..it was just like a small feeling of companionship, the lack of loneliness making the task a little less boring}''. 
At the same time, captions worked as careful reminders for participants, as P9 mentioned, ``\textit{The motivating part was actually more helpful. If I needed to see the mistake part, it felt like, am I doing a mistake too?}''.

% \subsubsection{Features for Enhancement}
% \subsubsection{Extending Body Double to Everyday Context}
% \subsubsection{Participants' behavior during VR sessions}
% interaction vs no interaction
\section{Discussion}

Our findings suggest that preferences for body doubles are not uniform but instead reflect a fundamental trade-off between ``motivational pressure'' and ``psychological safety''. 
Human doubles naturally provide stronger accountability, social comparison, and even competitiveness, all of which can motivate individuals to sustain performance. 
However, for participants sensitive to social pressure or prone to anxiety, this same realism was experienced as a source of discomfort, distraction, or stress. 
In contrast, AI doubles offered a ``safe accountability'' model, they kept the task moving and reduced loneliness without introducing the fear of being judged.
However, an important aspect to consider is that the degree of accountability experienced in human body doubling strongly depends on the nature of the relationship between the worker and the double.
When the body double is a friend or known peer, accountability may feel more supportive and collaborative and participants may interpret the friend’s presence as encouragement.
Prior HCI work on co-study groups and peer learning has shown that familiar peers can transform accountability from evaluative into cooperative.
On the other side, when the body double is an unknown person, accountability often takes on a more evaluative or competitive tone.
This evaluative dynamic can, for some, sharpen focus and effort as they strive to ``prove themselves'' in front of a stranger.
Considering broader applicability, this perspective suggests that body doubling may be most impactful when situated as a \textbf{configurable productivity aid} rather than a fixed methodology.
Employees with ADHD or social anxiety may prefer AI-based co-working tools that offer accountability without surveillance, while others may benefit from structured co-working with peers.
Our work demonstrates that AI body doubles can deliver performance benefits comparable to human doubles, while also offering the potential for personalized alignment between user needs and double type. 
This insight is significant for neurodivergent populations, offering new opportunities for designing scalable and accessible forms of body doubling support. 

\subsection{Implications for Design}

While our findings provide evidence that body doubling can be an effective productivity technique for adults with ADHD, this study also opens up broader questions about its role in future work contexts. In particular, our results point to opportunities for those who may try body doubling across different task types, under diverse platforms, and with AI doubles designed in varied ways—and potentially beyond these domains. This suggests the importance of adaptable, context-sensitive solutions rather than one-size-fits-all interventions. In the following, we outline four directions that future researchers can pursue: task-wise environments, device platforms and physical settings, AI body double design, and diverse brain types.

\vspace{-2mm}
\paragraph{\textbf{Task-Specific Body Doubling.}} \textit{How different task demands reshape the role of AI doubles}:

Future designs should carefully consider how task characteristics influence people’s needs for body doubling, though the ways to translate these needs into effective AI designs remain open.
(1) In high cognitive demand tasks such as creating, problem solving, or decision making, people may struggle with sustaining focus or managing complexity.
(2) In low cognitive demand or repetitive tasks, they may instead face monotony, boredom, or difficulty staying motivated.
(3) In physical tasks, such as exercise, pacing and sustained engagement can become central.
(4) Multitasking contexts raise challenges of coordination and divided attention.
(5) Under time pressure, stress and anxiety may dominate people’s experience.
(6) In environments full of distractions, the main concern may be the ability to maintain attention on the focal task.
These scenarios highlight diverse ways in which people’s cognitive or situational demands shape their reliance on body doubling, and raise open questions about how AI doubles can be designed to meet them.
These cases and examples show how varying task types place different demands on cognition and attention. By conducting more research on task-specific body doubling, we could scientifically characterize what forms of social support are most useful for people.

\vspace{-2mm}
\paragraph{\textbf{Understanding What Makes an AI a `Good' AI Body Double.}} \textit{Exploring presence, framing, and interaction strategies}:

One of the essential questions for future research is what aspects of AI’s interaction with humans can make it a ``good'' body double, since many design dimensions remain underexplored.
(1) The way body doubling is framed, for example, as a collaborative partnership versus a form of competition, might influence motivation in distinct ways.
(2) Conversation style is another open dimension, such as whether the AI should proactively initiate dialogue or remain passive.
(3) Relatedly, people may differ in how much they prefer the AI to explain its own progress while performing a task.
(4) Another uncertainty concerns whether the AI should provide proactive feedback about the user’s performance or wait until prompted.
(5) The visual presence of AI partners may also matter, as people could respond differently to abstract, cute, or realistic avatars.
(6) The baseline pace of the AI—whether fast, moderate, slow, or adaptive—could set different expectations and pressures for the user.
(7) Finally, methods for displaying the AI’s progress, ranging from real-time updates to simple visualizations, may shape how people interpret and align with the AI’s actions.
These cases and examples suggest that what makes an AI a ``good'' body double is far from settled, calling for systematic research to identify the characteristics that most effectively support people.

\vspace{-2mm}
\paragraph{\textbf{Body Doubling Across Diverse Minds, Conditions, and Ages.}} \textit{Tailoring doubles for neurodivergent and neurotypical workers, people with physical or cognitive disabilities, and across the lifespan}:

While body doubling is often studied in the context of ADHD, future research should investigate how different cognitive profiles, physical conditions, and age groups might shape its effectiveness.
(1) Among autistic adults, support needs may vary by DSM-5 level of required assistance, raising questions about how doubles can accommodate different intensities of cognitive or social support.
(2) Individuals with Asperger’s syndrome may have distinct preferences for communication style, predictability, and autonomy.
(3) Other intellectual and developmental disabilities (IDD) and neurodivergent profiles, such as dyslexia or Tourette’s, could also interact differently with the concept of body doubling.
(4) People with physical or sensory disabilities may require tailored forms of presence and interaction that account for accessibility.
(5) For children and adolescents, from toddlers to teenagers, the role of a body double may intersect with developmental stages of attention, self-regulation, and learning.
(6) Neurotypical adults may also benefit in different ways, such as using body doubling to manage distractions or improve accountability in knowledge work.
(7) Finally, seniors may engage with body doubling in the context of cognitive aging, memory challenges, or social isolation.
These cases and examples suggest that designing for diversity is not about a single model of body doubling, but about understanding how different minds, bodies, and ages shape what forms of social support are most meaningful.

\vspace{-2mm}
\paragraph{\textbf{Platform-Adaptive Body Doubling.}} \textit{Designing across VR, desktop, mobile, and wearables}:

Another consideration is how body doubling experiences may change depending on the platform through which they are delivered, since people’s activity contexts vary widely.
(1) On desktop environments, people are often sitting and engaged in cognitively demanding tasks, where sustained attention and structured interaction may be most relevant.
(2) On mobile devices, users may be moving between locations or engaging in shorter, fragmented sessions, raising questions about how body doubling can adapt to intermittent use and shifting contexts.
(3) On wearables such as smartwatches or fitness bands, people may be standing, walking, or exercising, where subtle cues and lightweight presence could become more meaningful than intensive interaction.
These cases and examples suggest that platform adaptation is not only about technical portability but also about aligning body doubling with the activity patterns and task types that fit each context.

As such, we have several factors, (1) human cognitive status–wise, (2) AI design and interaction–wise, (3) user demographic–wise, and (3) device and activity–wise, that remain largely unexplored in understanding body doubling.
These dimensions remind us that effectiveness is not a single phenomenon but a constellation of interacting conditions that shape how people experience support.
Advancing research in these directions could help us move from anecdotal or context-specific practices toward a more systematic, scientific characterization of body doubling.
Ultimately, by examining task demands, AI design choices, diverse populations, and platform contexts together, we can build a foundation for designing AI body doubles that are not only effective but also inclusive and adaptable across the complexities of everyday work and life.

\subsection{Limitations}
Our study has several limitations. 
The work primarily examined short-term effects, focusing on immediate performance and perceptions during VR sessions rather than long-term adoption, habit formation, or sustained productivity gains. 
We also evaluated only one instantiation of each body double (human and AI), without exploring variations in traits such as personality, communication style, or familiarity with the worker. 
While we extend this limitation in our implications, future work should systematically explore how these variations influence body doubling effectiveness.
Lastly, the work was conducted in a construction-inspired task setting, which provided ecological relevance for our focus but may limit generalizability to office-based contexts or other neurodiverse populations.
Broader investigations across diverse domains are necessary to confirm the applicability of AI and human body doubling beyond the construction context.
\section{Conclusion}

This work examined body doubling as a productivity support strategy for adults with ADHD in construction contexts.
In S1, we identified challenges and opportunities from workers' lived experiences, generating insights and design ideas for implementing body doubling with both humans and AI.
In S2, we evaluated its effects through a controlled experiment, comparing working alone with human and AI body doubles, and observing both the benefits of doubling and the subtle differences between human and AI presence.
Building on these findings, we outlined four implications for design that address task-specific demands, AI design considerations, the diversity of minds, conditions, and ages, and platform adaptation.

Our findings contribute to our understanding of how body doubling can be extended from an informal coping strategy into a structured, research-informed intervention.
More broadly, they suggest that body doubling offers a new lens on how social presence, whether human or AI, can be intentionally shaped to support focus, motivation, and productivity.
As AI and agentic systems grow more capable, the challenge will be to design them as partners that amplify people's ability to thrive under different conditions.
In this sense, body doubling may serve as an early example of how human–AI collaboration can be directed toward everyday struggles of attention and work, and ultimately toward creating more inclusive and empowering futures of productivity.

\bibliographystyle{ACM-Reference-Format}
\bibliography{ref}

\end{document}